\documentclass[]{aa}

\usepackage{graphicx}
\usepackage{natbib}
\usepackage{amsmath}
\usepackage{txfonts}
\bibpunct{(}{)}{;}{a}{}{,} % to follow the A&A style

\newcommand{\zs}{Z$_{\odot}$}
\newcommand{\ms}{M$_{\odot}$}

\newcommand{\lp}{Li$_{\rm P}$}

\begin{document}

   \title{The AMBRE project: a study of  Li evolution in the Galactic thin and thick discs  }

   \author{        
          N. Prantzos,
          \inst{1}
         P. de Laverny,
          \inst{2} 
          G. Guiglion,
          \inst{3}          
            A. Recio-Blanco,
          \inst{2} 
          C. C. Worley
           \inst{2,4}      
          }

   \institute{1. Institut d'Astrophysique de Paris, UMR7095 CNRS, Univ. P. \& M. Curie, 98bis Bd. Arago, 75104 Paris, France\\
      2. Université Côte d’Azur, Laboratoire Lagrange, Observatoire de la Côte d’Azur, CNRS, Blvd de l’Observatoire, CS 34229, 06304 Nice cedex 4, France \\
      3. Leibniz-Institut für Astrophysik Potsdam (AIP) An der Sternwarte 16, 14482 Potsdam \\
      4. Institute of Astronomy, University of Cambridge, Madingley Road, Cambridge CB3 0HA, UK }
              %\email{prantzos@iap.fr}              

   \date{Received ; accepted }

% \abstract{}{}{}{}{} vs
% 5 {} token are mandatory
 
  \abstract
   {Recent observations suggest a "double-branch" behaviour of Li/H versus metallicity in the local thick and thin discs. This is reminiscent of the corresponding O/Fe versus Fe/H behaviour, which has been explained as resulting from radial migration in the Milky Way disc. }
   {We study here the role of radial migration in shaping these observations. }
   {We use a semi-analytical model of disc evolution with updated chemical yields and parameterised radial migration. We explore the cases of long-lived (red giants of a few Gy lifetime) and shorter-lived (Asymptotic Giant Branch stars of several 10$^8$ yr)  stellar sources of Li, as well as those of low and high primordial Li. We show that both factors play a key role in the overall Li evolution.}
   {We find that the observed "two-branch" Li behaviour is only directly obtained in the case of long-lived stellar  Li sources and low primordial Li. In all other cases,
   the data imply  systematic Li depletion in stellar envelopes, thus no simple picture of the Li evolution can be obtained. This concerns also the reported Li/H decrease at supersolar metallicities.}
   {}


\titlerunning{Li evolution in the Galactic thin and thick discs}
   \maketitle

\section{Introduction}
\label{sec:Intro}

Li is known to have the most complex behaviour among all chemical elements, regarding its nucleosynthesis and chemical evolution. It is the only one to have three confirmed nucleosynthetic sites: primordial, cosmic rays and stellar,  the latter producing the majority of Li. It is unknown yet whether the stellar sources are novae (through explosive H-burning), red giants (shell H-burning), Asymptotic Giant Branch (AGB) stars (hot-bottom burning) or core-collapse supernova (CCSN, through $\nu$-induced nucleosynthesis). \cite{Prantzos2012b}, using detailed models for the chemical evolution of all light elements,   showed that published yields from the aforementioned sources fail to reproduce the solar Li abundance. However, the recent  reports of the presence of substantial Li amounts in novae ejecta  \citep{Izzo2015,Tajitsu2015,Tajitsu2016}, have revived interest in the role of novae as Li producers (Sect. \ref{sub:Model2}).
 
From an observational point of view, the discovery of the Li "plateau" in halo stars of low metallicity confirmed that Li is indeed produced in the Big Bang \citep{Spite1982}. The so-called "Spite plateau" displays a remarquably constant value of Li/H with metallicity [Fe/H], during the whole evolution of the Galactic halo, that is up to [Fe/H]\footnote{Throughout the text we use standard spectroscopic notation: \\
- $ \rm [Fe/H] = log_{10}(N_{Fe}/N_H)-log_{10}(N_{Fe}/N_H)_{\odot}$ \\
- $\rm A(Li) = log_{10}(N_{Li}/N_H) + 12$ \\
where $\rm N$ are abundances by number.}$\sim$-1; this corresponds to about 1 Gy of evolution, this timescale being estimated from the inferred dispersion  in the ages of globular clusters. 

Subsequent refinements in calculations of Standard Big Bang Nucleosynthesis (SBBN in the following) revealed a discrepancy between theory and observations: the observed Spite plateau, long thought to reflect the primordial Li abundance (\lp, \ hereafter), lies by a factor of $\sim$3-4 below the one  found in SBBN calculations that corresponds to the cosmic baryonic density. The latter is  inferred from observations of the  Cosmic Microwave Background (CMB) and from the D/H value observed in high-redshift gas clouds \citep{Cooke2016}. Despite considerable efforts, no satisfactory solution to that problem exists yet \citep{Coc2016}. 

At metallicities [Fe/H]$>$-1, corresponding to the evolution of the Milky Way disc, 
Li/H rises steadily from the halo plateau up to its solar value (for example,  \cite{Lambert2004} and references therein). This corresponds to an increase by a factor of 
$\sim$10 if \lp \ is given by the Spite plateau or a factor of $\sim$3 if \lp \ is the one of SBBN. For a long time it was thought that this sharp rise pointed to a delayed Li source, operating essentially after the first Gy of galactic evolution. Low-mass red giant stars (1-2 \ms) and, perhaps, novae  naturally fulfill this requirement. In the case of other, shorter-lived,  candidate sources,  (AGB stars  of mass M$>$2 \ms \ or core collapse supernovae) the resulting Li abundance breaks the Spite plateau earlier than [Fe/H]$\sim$-1 and one needs then to invoke systematic depletion of Li in the stellar envelopes in order to explain the data.

In the last few years, observations of Li in the thick and thin discs of the Milky Way (MW) have added a new twist to the Li saga. 
\cite{Ramirez2012}  found that Li/H remains  approximately constant with [Fe/H] during the evolution of the thick disc, at about its halo plateau value.  On the other hand, 
\cite{Guiglion2016}  find a slight increase of Li/H during the evolution of the thick disc. The results of those studies are compatible with each other, within error bars: stars of the thick disc, although at metallicities [Fe/H]$>$-1, show  little  rise in their Li/H  abundance. On the other hand, all studies find that the sharp rise of Li/H previously thought to characterise the local Galactic disc, concerns in fact {\it only the thin disc}, thought to be 7-9 Gy old.

The observed "two-branch" behaviour of the Li/H evolution in the thin and thick discs is reminiscent of the corresponding behaviour of $\alpha$/Fe versus Fe/H: stars of the same metallicity have a higher $\alpha$/Fe ratio if they belong to the thick disc than to the thin disc; see, for example, \citet{Adibekyan2012,Recio2014}. This behaviour can be interpreted in terms of radial migration \citep{SB2009}: because of the inside-out formation, the inner disc evolves within a shorter  timescale than the outer one and is mostly enriched by massive star explosions (CCSN)  and not by thermonuclear supernovae (SNIa); thus, its stars have a high $\alpha$/Fe ratio\footnote{Massive stars evolve on short timescales (a few My) and produce all oxygen in the galaxy, but only part of its Fe content ($\sim$1/3 to 1/2), thus having a supersolar ratio [O/Fe]$ \sim$0.3-0.5 in their ejecta. The remaining Fe is produced on longer time scales (100 My to several Gy) by SNIa and the overall [O/Fe] ratio progressively decreases.}. Some of them migrate to the outer disc and constitute a $\alpha$/Fe-rich population, while stars formed further out (e;g. in the Solar neighbourhood) constitute a $\alpha$/Fe-poor population, being formed on longer timescales and enriched also by SNIa (see also  \cite{Minchev2013,Kubryk2015a,Kubryk2015b}).

In this work, we study the implications of radial migration on the abundance patterns of Li/H in the thin and thick discs, based on the model of \cite{Kubryk2015a}. We account for the uncertainties currently affecting Li production and evolution by studying the cases of long-lived (red giants) and shorter lived (AGBs) Li sources, as well as a high (SBBN) and a low ("Spite plateau") value of primordial Li (\lp). The plan of the paper is as follows: in Sect. \ref{sec:Observations} we present the relevant observational data in some detail; in Sect. \ref{sec:Model} we discuss the difficulties in modelling Li evolution and we present the adopted model, while in Sects. \ref{subsec:Results1} and \ref{subsec:Results2} we present and discuss our results, which are summarized in Sect. \ref{sec:Summary}.

\section{Observational constraints on Li evolution in the Milky Way}    
\label{sec:Observations}

Observational data characterising the evolution of Li in the local (thin and thick) discs of   MW are displayed in  Fig 1. Corresponding data for the halo, also known as the "Spite plateau", appear in a schematic way on the left of that Figure, where the high primordial Li abundance of SBBN is also diplayed.

The difference of $\sim$0.5 dex  separating  the SBBN value from the Spite plateau remains a puzzle today and  well summarized in several recent reviews, for example, \cite{Coc2016} and references therein. If the true primordial Li is the one of SBBN, depletion before the formation of the halo stars or inside them during the subsequent 12 Gy of evolution
must have reduced it in an extremely uniform way to the Spite plateau; no satisfactory mechanism for the required uniform depletion over a range of $\sim$2 dex in [Fe/H] has been found yet (see for example, 
the recent study of \cite{Gruyters2016} -and references therein - for the case of atomic diffusion). On the other hand, if the true primordial Li is the one of the Spite plateau, non-standard physical processes during  the primordial nucleosynthesis should have shaped it; again, there is no agreement on  such a process today. For those reasons {\it we shall consider in this work both cases, starting our calculations with either a high or a low primordial \lp }.

The observational study of the Li behaviour in the Milky Way disc requires
the analysis of statistically significant and homogeneous samples and should
cover a large range of metallicities.
There are in the literature very few studies reporting
such a Li analysis for at least few hundreds of stars, for example, \cite{Lambert2004,Ramirez2012,Delgado2015}. 
The one of \cite{Guiglion2016} (hereafter G16) investigates  the Galactic Li behaviour for a much larger sample of about 7,300 stars; this latter study is based on the homogeneous analysis of ESO high-resolution
archive spectra analyzed in the AMBRE Project \citep{Laverny2013}.

\begin{figure}
\begin{center}
\includegraphics[width=0.49\textwidth]{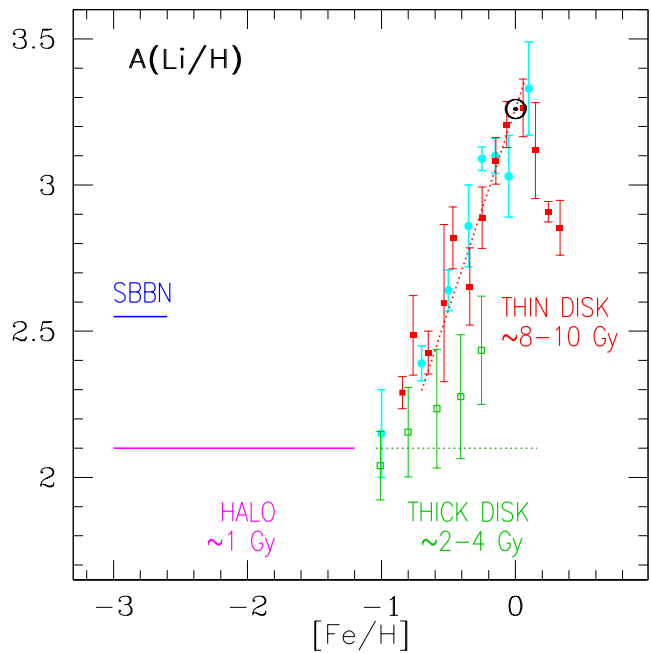}
\caption[]{Observational constraints to the evolution of Li in stars of the halo and the thick and thin discs of the Milky Way. The Li abundance resulting from SBBN is shown in the upper left part of the Figure and the "Spite plateau" for halo stars on the bottom left. Corresponding upper limits for the thick (green) and thin (red) discs are shown  on the right. The dotted lines  are upper limits from \cite{Ramirez2012}.  Blue filled circles with error bars represent the average Li abundances in the most Li rich stars of the corresponding metallicity bin in the sample of \cite{Lambert2004}; no distinction between thin and thick discs is made in that work. The squares (filled for the thin disc and open for the thick disc) with error bars represent the average Li/H abundances of the most Li-rich stars in the AMBRE data of \cite{Guiglion2016}. 
}
\label{Fig:LiTime}
\end{center}
\end{figure}

From the derived Li abundances in stellar atmospheres,
most of these works estimated the ISM Li abundance behaviour versus metallicity
by considering the upper envelope of the Li distribution, a good indicator
of the ISM initial abundances before any Li depletion in stars. 
We, however, point out that all these studies are focussed on the Solar neighbourhood.
There is still a lack of a wide survey of Li abundances in more distant
regions of the Milky Way, well beyond $\sim$500~pc from the Sun.
Moreover, thanks to additional chemical and/or kinematical information,
the above samples have been able to disentangle the Li behaviour
in the thin and the thick discs.

Following a now commonly adopted approach in Galactic spectroscopic surveys, for example, \cite{Recio2014},
G16 used a purely chemical criterium, that is, the [$\alpha$/Fe] content of the AMBRE stars 
in order to tag them as  thin ($\alpha$-poor) or thick ($\alpha$-rich) disc members.
In contrast, \cite{Ramirez2012}  adopted a procedure based on kinematic criteria, 
available because of
already known Galactic space velocities for all their sample stars whereas 
\cite{Delgado2015} adopted both kinematics  - actually the ones of \cite{Ramirez2012} -  and chemistry
to disentangle the two  discs. Finally, the \cite{Lambert2004} sample
is almost completely dominated by Solar vicinity thin disc stars, several of them having 
accurate Hipparcos parallaxes and circular orbits. We note that these different methodologies
adopted to label thin/thick stars
could explain some of the small differences found between these studies
for the Li behaviour in the Galaxy.
We wish to point out, however, that we tend to favour the chemical criterium
to tag the thin/thick discs. The reason is that  the classical kinematically-based definition
could blur our vision of the two discs because of potential overlapping in the
kinematical distributions of both discs \citep{Bovy2012}. Moreover,
the chemical definition appears to be  a better approach than the kinematical one
since chemical abundances are formally independent of the disc structure, even if they may  correlate with it 
(see discussion in \cite{Recio2014}).

Regarding the evolution of Li in the thin disc shown in Fig.1, 
there is  good agreement between
all these studies until the formation of
the Sun: a steady increase from slightly above the Spite plateau
up to the Solar System value. The reported Li increase is a bit larger than 
1~dex between [Fe/H]= -1.0 and +0.0~dex. 

On the other hand, Li abundances in the local thick disc 
are clearly lower than those of the thin disc although
some differences are reported by these studies.
G16 found a slow increase of Li abundances for (chemically defined) thick disc stars
by about 0.4~dex, starting from the Spite plateau. In contrast,
\cite{Ramirez2012} reported a rather flat Li versus Fe behaviour
up to solar metallicity \zs \ (although almost no thick disc stars are expected
at such high metallicity), as an extension of the Spite plateau.
We note, however, that these two patterns are compatible with each other  within 2$\sigma$.
These findings are in disagreement with the decrease of Li with metallicity reported by 
\cite{Delgado2015} for their most metal-rich thick disc stars,
but small statistics and lack of Li-rich stars could explain this strange behaviour.

Our last remark concerns the behaviour of Li at supersolar
metallicities in the thin disc: G16 found that the upper envelope of
the AMBRE data decreases with metallicity by about 0.5~dex up to [Fe/H]
$\sim$0.4~dex. This phenomenon was already suspected by \cite{Delgado2015}.
If this is confirmed, it would be the first time that the abundance of an element 
produced by stars during Galactic evolution is found to be decreasing
in the past few Gy. We shall consider this unexpected behaviour
at the end of Sect. 4, but without discussing in detail its physical implications.

\section{Model of Li evolution }
\label{sec:Model}

\subsection{Difficulties in modelling Li evolution }
\label{sub:Model1}

Modelling the evolution of Li in the thin and thick disc of the MW is hampered by a number of  difficulties.

1. The origin of the thick disc and its relation to the thin disc are not yet clearly identified, see for example, the discussion in \cite{Kawata2016}. In those conditions, it is hard to provide a universally accepted framework for the evolution of Li. In this work, we shall adopt the  model of \cite{Kubryk2015a} with parametrized radial migration which reproduces satisfactorily several of the observed chemical properties of the MW thin and thick discs. 

2. As emphasised in the Introduction, the main source of  solar Li is not yet known. In fact, Li - and, in particular, its heavy isotope $^7$Li - is the only element to have more than two confirmed types of sources: primordial nucleosynthesis, spallation reactions in cosmic rays  and stellar production \citep{Prantzos1993,Abia1995,Romano2001,Travaglio2001,Prantzos2012b}.  The first two are reasonably well constrained from observations and they appear unable to provide more than 30 \% of solar Li, see \cite{Prantzos2012b} for a recent study. The latter Li source is unknown at present: novae, red giants, AGB and neutrino-nucleosynthesis in CCSN have been suggested over the years, but the current estimates of the yields of all those sources appear substantially lower than the observational requirement, - by a factor of ten \citep{Prantzos2012b}. The AMBRE observations provide a new and important constraint for the stellar Li sources, since the thick disc data correspond to the early evolution of the MW disc, that is, its first 3-4 Gy: Li appears to evolve little during that early period, thus constraining the timescale of its stellar source (see below).

3. The current uncertainty on the primordial Li abundance \lp \ (Sect. \ref{sec:Observations})  makes it impossible to adopt a well determined value for the initial Li abundance in the model. Since the value of \lp \ strongly affects the results of the chemical evolution calculations,    we shall consider  two cases here: one with \lp \  given from SBBN (high \lp) and another based on the observed value of the Spite plateau (low \lp).

4. Usually, the observed abundances in the surfaces of low-mass stars  reflect
the corresponding abundances of the gas from which those stars formed billions of years ago and they are used to constrain the chemical evolution models. In contrast, Li is a fragile element burning at a temperature T$>$2 MK which is encountered in the base of the convective envelope of all but the hottest main sequence stars. This is the case for  the Sun, the photospheric Li abundance of which is  approximately 200 times lower than the pre-solar Li abundance obtained by meteoritic data. In those conditions, a comparison between observations and model results is not straightforward (\cite{Lyubimkov2016} and references therein). A way out of this difficulty is to adopt the upper envelope of the corresponding Li observations as the true gas Li abundance at each time (or metallicity);  for example, \cite{Lambert2004}. This assumption brings observations closer to the Li abundance in the gas forming those stars. However, it suffers from a potential drawback, related to the current discrepancy between the observed Spite plateau  and the results of SBBN, as discussed in Sect. \ref{sec:Observations}: if that discrepancy is real and it is due
to depletion in the envelopes of halo stars, then such depletion may also affect stars in the disc, making a direct comparison of model results to observations impossible. In that case, comparison of model to observations may only provide some hints for the stellar depletion. We shall see that this is indeed the case when a high \lp \ value is adopted or short-lived sources of Li are assumed.

\subsection{The model}
\label{sub:Model2}

We summarize here the main features of the \cite{Kubryk2015a} model.  
The Galactic disc is gradually built up by infall of primordial gas in the potential well of a
dark matter halo with mass of 10$^{12}$ \ms \ whose evolution is obtained from
numerical simulations. The star formation rate depends on the local surface
density of molecular gas. The  infall time-scales are shorter in the inner regions, while they
increase outwards reaching 7 Gyr at 7 kpc. 
The model takes into account the radial flows of gas driven by a bar formed 6 Gyr ago which pushes
gas inwards and outwards of the corotation. 

Stars move radially due to  epicyclic motions
(blurring) and  variation in their guiding radius (churning).
The innovative aspect of the model is  that it accounts for the fact that radial migration moves around
not only "passive tracers" of chemical evolution (that is, long-lived stars, keeping in
their photospheres the chemical composition of the gas at the time and place of their birth), but
also "active agents" of chemical evolution, i. e., long-lived nucleosynthesis sources
such as SNIa producing Fe and low-mass stars producing s-process elements and, perhaps, Li.
In the \cite{Kubryk2015a,Kubryk2015b} model, the thick disc is assumed to be the oldest part of the Galactic disc, older than $\sim$9 Gy (see also \cite{Binney2014}) , while the thin disc is younger that 9 Gy.

The \cite{Kubryk2015a,Kubryk2015b} version of the model uses  the metallicity-dependent yields of \cite{Nomoto2013}.
In the version used in the present work we adopt the new metallicity-dependent
yields by Chieffi \& Limongi (2017, paper in preparation) which include the effect of mass loss and  stellar rotation and cover the whole range of massive stars, up to 120 \ms.
 A phenomenological  rate of SNIa is adopted,  based on observations
of extragalactic SNIa, while their  metallicity-dependent yields are from \cite{Iwamoto1999}. The initial mass
function (IMF) of \cite{Kroupa2002} with a slope 1.5 for the high masses, is adopted.

In  order  to calculate the rate of ejecta (both for stars and SNIa) as a function of time,
the    formalism of single particle population is used
 because it can account for  the radial displacements of
nucleosynthesis sources and in particular of SNIa as discussed in Appendix C in \cite{Kubryk2015a}. This is also very important in our case, since some of the suggested stellar Li sources are low-mass, long-lived stars. 

We calculate the Li evolution with the detailed treatment presented in \cite{Prantzos2012b}. Its production by galactic cosmic rays is calculated in a fully self-consistent way, taking into account the energetics of supernovae and the composition of the ISM and of stellar winds. As discussed extensively in recent work, there are strong indices that cosmic rays are accelerated from  - and have the composition of - stellar wind material, both on empirical   and on observational grounds; the former concerns the observed Ne22/Ne20 ratio on GCR \citep{Prantzos2012a}, while the latter is related to the observed linear behaviour of purely spallogenic Be with Fe \citep{Prantzos2012b}.

As already mentioned, the stellar sources of Li are unknown at present, despite the larger number of candidates, larger than for any other element: novae, red giants, AGB stars and CCSN. The Li yields of all those sources are  uncertain and their dependence on metallicity poorly understood. In the case of single stellar sources, though, the time scale of their evolution is well known, once the stellar IMF and star formation rate are fixed by the chemical evolution model. For our purpose, we shall consider here only two types of source:  long-lived ones (evolving on timescales larger than the halo phase, that is, $\sim$Gy) and shorter-lived ones (less than a Gy). In doing that, we are motivated by the fact that the observed double-branch behaviour of [O/Fe] can be interpreted in terms of a differential evolution of the main sources of O (CCSN, evolving on a timescale of a few My) and Fe (SNIa, evolving on a $\sim$Gy timescale). 

In an analogous way,  as representative long-lived sources of Li, here we study  red giant stars with masses  in the range 1.2 - 1.5 \ms, that is, with lifetimes larger than 2  Gy. We shall show that in this case and under the assumption of low  \lp, \ the observational constraints of Fig. 1 can be reproduced for the thin and thick disc with no need for internal Li depletion in stellar envelopes. We shall also show that all other cases (starting with high \lp \ and/or adopting shorter-lived Li sources) require substantial internal Li depletion in order to explain the data. Finally, we shall explore the possibility of a metallicity-dependent Li yields (reduced at supersolar metallicities) in order to explain the putative decrease of Li revealed by AMBRE at \zs $>$0 \citep{Guiglion2016}. 

Despite the revived interest in novae after the recent reports of Li detection in three such systems 
\citep{Izzo2015,Tajitsu2015,Tajitsu2016}, we shall not explicitly consider those sources here. The reason is twofold.
First, although the present rate of novae in the Galaxy is known, the
past one is not, and it is impossible to infer it from any theoretical argument\footnote{One might be tempted
to attach the nova rate to the SNIa rate, since both systems involve white dwarfs in binaries; however,  in SNIa the white dwarf explodes only
once and disappears, while novae are recurrent and the timescale of their repetitive explosions is not known}. In those conditions, "playing" with the evolution of
the nova rate in order to probe the resulting Li evolution is a hopeless game. The second reason is related to a  a well known, but rarely mentioned, problem with nova nucleosynthesis. In all nova
models, even if  Li is  made in substantial amounts, there is invariably an even more important overproduction of the
minor CNO isotopes ($^{13}$C, $^{15}$N and $^{17}$O), for example, \cite{Jose1998,Denissenkov2014}. This is valid both for CO and ONe novae and is a serious handicap for novae as major Li sources: if one assumes that their theoretical yields are  approximately constant during galactic evolution, then the minor CNO isotopes would be found largely overproduced, as noticed in \cite{Jose1998}. These considerations do not imply that novae should be definitely excluded as potential Li sources. It may be that, up to now overlooked effects - due for example, to the 3D structure of nova explosions, not considered so far in models of detailed nova nucleosynthesis - help to alleviate the problem of the overproduction of minor CNO isotopes. However,   at the present stage of our knowledge it is difficult to make conclusions about the contribution of those objects to the galactic evolution of Li. Unfortunately, this is also true for other Li candidate sources, as we  discuss in the following.

\section{Results and Dicsussion}
\label{sec:Results}

As extensively discussed in \cite{Kubryk2015a}, the adopted model reproduces a large number of observational features of the MW, concerning both gas and stars. In particular, it reproduces the present day profiles of atomic and molecular  gas, star formation rate (SFR) and gaseous chemical abundances for key elements, like O and Fe. It also reproduces the total masses and the stellar profiles of the chemically defined thin and thick discs. In the solar neighbourhood, it  satisfactorily  reproduces the age-[Fe/H] relationship and its dispersion,  the metallicity distribution and the "two-branch" behaviour  of [$\alpha$/Fe] versus [Fe/H] in the local thin and thick discs. It should be noted that, in \cite{Kubryk2015a} the thick disc is defined as the oldest part of the MW disc, formed in the first 3 Gyr mostly in the inner regions; this timescale plays an important role in the evolution of Li in this model.

\subsection{Li from a long-lived source}
\label{subsec:Results1}

\begin{figure}
\begin{center}
\includegraphics[,width=0.49\textwidth]{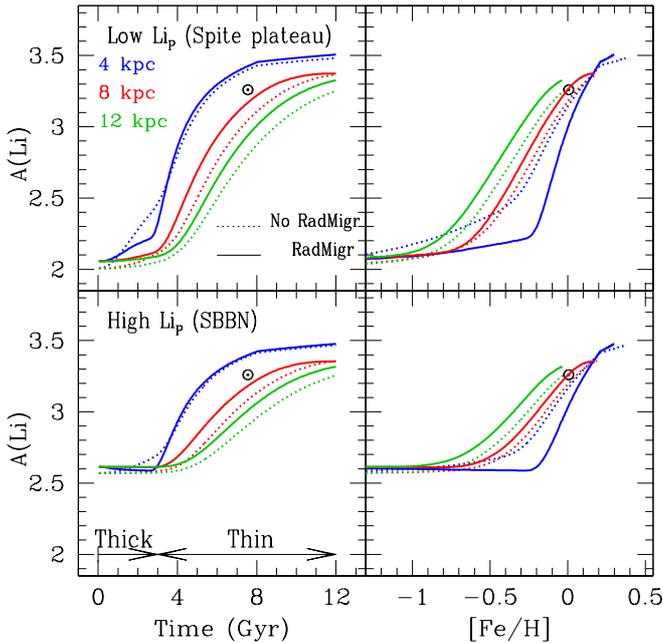}
\caption[]{Evolution of the Li abundance in the gas for three radial zones, at 4 (blue), 8 (red) and 12 (green) kpc from the Galactic centre, respectively. {\it Dotted curves} show the situation if radial migration were ignored, whereas {\it solid curves} display the results with radial migration. {\it Top panels:} Low primordial Li assumed (Spite plateau); {\it Bottom panels:} high primordial Li assumed (SBBN).  {\it Left:} Li abundance as a function of time. {\it Right}: Li abundance as a function of [Fe/H]. The stellar Li source is low-mass red giants.
} 
\label{Fig:LiFeH}
\end{center}
\end{figure}

In the panels of Fig. \ref{Fig:LiFeH} we present our results for the evolution of 
Li/H, assuming its stellar source is low-mass red giants.  The production of Li in that source through "cool-bottom burning"  has been extensively investigated by \cite{Sackmann1999}, who did not provide Li yields but emphasised the many uncertainties in the Li production versus destruction mechanisms. Similarly, the recent study of \cite{Lattanzio2015} underlines the high sensitivity of Li abundance in red giants to the physical conditions, and in particular, to the numerical treatment of thermohaline mixing. In view of these uncertainties, we  adopt here a Li  yield of 10$^{-7}$ \ms \ for those stars in order to reproduce the solar Li abundance for an average star 4.5 Gy old and present in the solar neighbourhood today. We consider the cases of low primordial \lp \  (top panels) and high \lp \ (bottom panels), as a function of time (left) and of [Fe/H] (right). The panels display the evolution of Li/H in the gas for 3 radial zones, at galactocentric distances of 4, 8 and 12 kpc, respectively. For each zone, we show the results of in-situ chemical evolution, that is, gas abundance in the absence of any radial migration (dotted curves) and with radial migration included (solid curves). The latter differs from the former in that some of the long-lived sources of Li have time to move away from their birth place before releasing it into the ISM.

In all cases, Li/H starts increasing substantially, that is, getting above its \lp \ value, after a time scale of $\sim$2 Gy at 4 kpc, $\sim$3 Gy at 8 kpc and $\sim$4 Gy at 12 kpc. The reason for those different timescales, despite the fact that the stellar source is always the same, is that the vigourous early star formation in the inner disc  allows Li to "break"  the plateau of  its primordial value earlier in that zone; in contrast, there is little star formation in the outer disc early on (because of the inside-out disc formation) and thus more time is required for Li to rise above \lp. 

The impact of radial migration is easier to understand when the results are plotted as function of time (left panels in Fig. \ref{Fig:LiFeH}): at each radius, Li/H is higher at a given time when radial migration is included. The reason is that low-mass stars migrate from the inner disc into those zones  and enrich them with Li. There is net Li enrichment with respect to the case of no radial migration, because the migrating stars   are more numerous than the stars born in situ which left those same zones; see also Fig. 6 in \cite{Kubryk2015a}.

It is not straightforward  to understand the same results plotted as function of Fe/H (right panels in Fig. \ref{Fig:LiFeH}). Now, Li/H is again higher with radial migration in the zones at 8 and 12 kpc, but not at 4 kpc. The reason for this  behaviour is the competition between the timescales for radial migration on the one hand and those of the Li {\it and} Fe sources on the other: at 4 kpc, SNIa enrich rapidly the gas to high Fe/H values, within a few 10$^8$ yr, before low-mass stars enrich it with Li. When Li enrichment above the primordial plateau starts (after  $\sim$1 Gy,   a fraction of the Li producing stars born at 4 kpc has already migrated and the resulting Li enrichment in that zone is smaller than without radial migration.  In contrast, at 8 kpc or further outwards the opposite effect occurs (migrating sources from the inner disc to those regions outnumber local migrating  sources, as explained in the previous paragraph)  resulting in a Li enrichment which is larger than without radial migration for a given metallicity.

In summary, the evolution of Li, or of any other element produced by long-lived sources (with timescales of $>$1 Gy), is much more complex in the presence of radial migration and depends considerably on the magnitude of that effect. Fe is less affected, because many  SNIa explode within the first Gy after the formation of the binary system, and suffer little radial migration in that interval.

An inspection of the left and right panels of Fig. \ref{Fig:LiFeH} shows that a) the abundance of Li/H never exceeds a maximal value of A(Li)=2.5 (for low \lp) or A(Li)=2.6 (for high \lp) in the first 3 Gyr, that is, up to  the end of the thick disc phase in the framework of the adopted model, and b) that maximal value  is reached at [Fe/H]$\sim$-0.1 in the inner zones, that is,  those zones 
have quasi-solar Fe but a few times less than solar Li towards the end of the thick disc phase.
Obviously, if some of the stars of those inner zones migrate to the local disc, one can easily understand the presence of stars belonging to the branch of "high Fe/H and low Li/H", as seen in the observations of AMBRE.

\begin{figure}
\begin{center}
\includegraphics[width=0.49\textwidth]{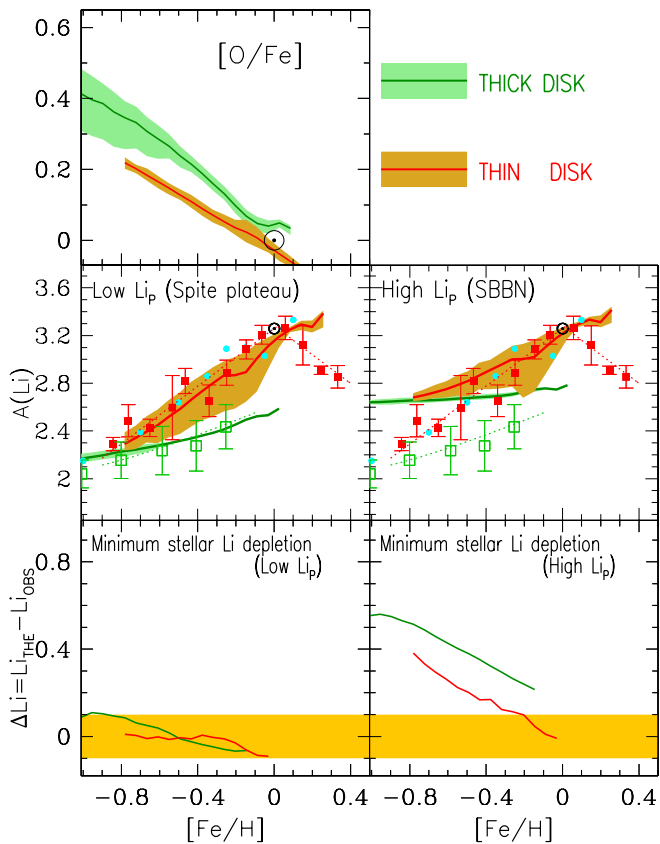}
\caption[]{Evolution of abundances in the thick (green curve) and thin (red curve) disc of the Milky Way.  Shaded areas display corresponding 1 $\sigma$ dispersion. The stellar Li source is 1.2-1.6 \ms \ red giants. {\it Top} : Evolution of O/Fe. {\it Middle}: Evolution of Li/H, starting with low ({\it left}) or high ({\it right}) primordial Li. Observations are from Fig. 1 and the dotted lines are fits to those data, used as approximations to derive the minimum depletion of Li (see text). {\it Bottom}: Corresponding evolution of the required level of minimum depletion of Li  inside stars.  Shaded aereas in the bottom figure correspond to 1$\sigma$ errors of 0.1 dex.
}
\label{Fig:LiversusFeH2}
\end{center}
\end{figure}

Fig. \ref{Fig:LiversusFeH2} shows clearly that result. Its upper panel  shows [O/Fe] versus [Fe/H] in the local disc, displaying the classical "two-branch" behaviour for the thin and thick disc. The latter (green curve) has old stars of higher O/Fe for a given Fe/H, because they come mostly from the rapidly evolving innermost regions: they have rapidly been enriched to high metallicities in O and Fe by CCSN, before SNIa produce their Fe and cause the decline in their O/Fe ratios. In contrast, stars of the local thin disc are younger and mostly local: they were formed in regions that reached high Fe/H later, when SNIa had substantially reduced the O/Fe ratio of their gas.
This "two-branch" behaviour is the same as in \cite{Kubryk2015a}, albeit slightly different because of the use of new massive star yields in the present work (see Sect. \ref{sec:Model}).

The corresponding evolution of Li is displayed in the middle panels of Fig. \ref{Fig:LiversusFeH2}, starting with low \lp (left) or high \lp (right). A "two-branch" behaviour is obtained in both cases, as expected after the analysis of Fig. \ref{Fig:LiFeH}. In the former case (low \lp), the results for the thin and thick discs reproduce nicely the observations of AMBRE for [Fe/H]$<$0.1 dex. In the latter (high \lp) Li/H in the thick disc is flat, since the original Li/H abundance is too high and Li production by low mass stars cannot produce enough Li to "break" the plateau in the first 3 Gy, even in the innermost zones.

In view of those results, one may be tempted to conclude that the analysis of both the halo data (Sect. \ref{sec:Observations}) and the thin/thick disc data (previous paragraphs) favours a low \lp \ (Spite plateau) and not the one of SBBN. This conclusion offers the simplest solution to the primordial Li puzzle from the point of view of stellar and galactic evolution. In the following we define the {\it minimum depletion} inside stellar envelopes as the maximum amount of Li in the stars brought to the solar neighbourhood (as predicted by our model) minus the upper envelope of Li data (as observed in thin and thick discs), that is,   $D_{min}$=Li$_{model}$-Li$_{Up,obs}$, where all abundances are functions of [Fe/H]. For Li$_{model}$ we adopt not the average values (curves of Fig. \ref{Fig:LiversusFeH2}) but the maximum ones (upper envelope of green and gold shaded regions), that is, the upper envelope of the 1$\sigma$ dispersion in the model results.  

This quantity is plotted in the bottom  panels of Fig. \ref{Fig:LiversusFeH2}. In the case of low \lp \ (left panel)  there is no need to assume any Li depletion in the envelopes of the hottest observed stars. The model  curve  closely follows the upper envelope of the data and the required minimum depletion is lower than the error in the data (typically of 0.1 dex, within the shaded region).

\begin{figure}
\begin{center}
\includegraphics[width=0.49\textwidth]{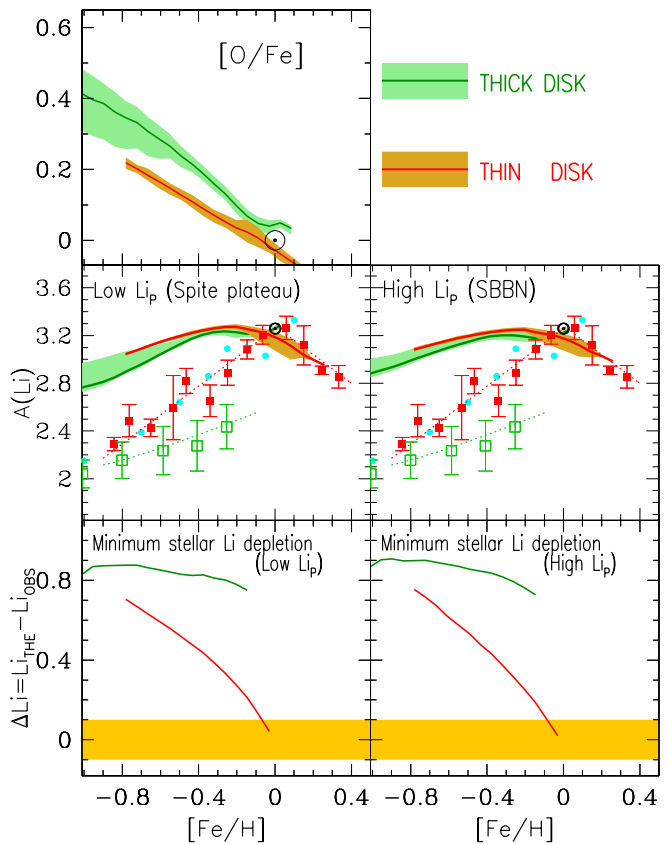}
\caption[]{Same as Fig. \ref{Fig:LiversusFeH2}, but this time the stellar Li source is assumed to be 2-5 \ms \ AGB stars {\it and} their Li yields are assumed to decrease with metallicity above [Fe/H]=-0.2, in order to reproduce the Li/H decrease obtained by AMBRE (see text). 
}
\label{Fig:LiversusFeH_high}
\end{center}
\end{figure}

In contrast, the bottom right-hand panel displays the implications for stellar depletion if \lp \ is the one of SBBN.  At each metallicity,  minimum depletion  is more important for the old stars of the thick disc than for the stars of the thin disc. Its highest value ($\sim$0.5 dex or a factor of 3) is obtained for the oldest and most metal-poor stars of the thick disc and is comparable to the depletion required to understand the Spite plateau if the \lp \ value of SBBN is correct. Thus, if the primordial Li abundance is as high as the one of SBBN,  it becomes impossible to constrain the  subsequent Li evolution - and its stellar source - through observations of disc stars; such observations may only constrain a combination of Li stellar sources and Li depletion in low-mass stars. We show here that, in the framework of our disc model and assuming low-mass stars as the stellar Li source,  its {\it minimum} stellar depletion $D_{min}$ has to be a declining function of  both metallicity and age, since stars of the thin disc having the same metallicity as the ones of the thick disc display  smaller Li depletion. Whether such a dependence can be explained by stellar models is beyond the scope of this work.
We simply notice here that this is the case in some stellar models invoquing atomic diffusion to explain the discrepancy between the Spite plateau and \lp \ from SBBN (for example, \cite{Gruyters2016} and references therein).

\subsection{Li from a short-lived source}
\label{subsec:Results2}

In Fig. \ref{Fig:LiversusFeH_high} we assume that the stellar Li source is AGB stars in the $\sim$2-5 \ms \ range with lifetimes of the order of  10$^8$-10$^9$ yr. Li in those stars can be produced in the AGB phase through "hot-bottom burning" (HBB) and the Fowler-Cameron mechanism. The amount of Li produced in that stage of stellar evolution depends a lot on the physical conditions, in particular the treatment of mixing and the mass loss rate of the star. This is illustrated in the recent works of \cite{Karakas2016} for stars in the 1-7.5 \ms \ range (which display HBB above 4 \ms) and \cite{Doherty2014} for more massive stars (6.5-9 \ms) reaching the super-AGB phase. In the former case, no Li production is found at metallicity \zs, a little  Li is produced  around 4 \ms \ at 1/2 \zs \ and important production occurs only at 2 \zs. In the  study of \cite{Doherty2014}, some Li is produced only in the most massive stars, around 8 \ms. In view of the results obtained in \cite{Prantzos2012b}, we conclude that in neither case the Li yields are sufficient to reproduce the solar Li abundance, by at least a factor of two. This is not a bad performance, but the uncertainties preclude any firm conclusion.

In view of the uncertainties in the Li yields, and since our purpose is merely to test the evolutionary timescale of the putative Li source, we adopt here for illustration purposes a Li yield of 2 10$^{-7}$ \ms \ in the mass range 2-5 \ms. We find then that there is no clear "two-branch" behaviour for Li. The reason is that Li sources evolve in the whole disc almost in step with - or even faster than -  the main Fe sources (SNIa), which span a similar range of progenitor lifetimes. Li/H versus Fe/H evolution is then the same everywhere (in contrast to the situation in Fig. \ref{Fig:LiFeH}); in those conditions, old stars migrating in the solar vicinity from the inner disc do not differ substantially in their Li versus Fe behaviour from locally born stars: for a given [Fe/H], they have similar  Li/Fe ratios. As a result, the evolutionary tracks of the thin and thick discs in the Li/H versus Fe/H plane are quite similar, giving the impression of a "single-branch" behaviour. Observations appear of no help  in  disentangling Li sources in that case. However, they  imply a much larger Li depletion in the thick than in the thin disc (bottom panels of Fig. \ref{Fig:LiversusFeH_high}), while the difference was quite small in the case of low-mass Li sources (bottom panels of Fig. \ref{Fig:LiversusFeH2}). We are unaware of any plausible explanation for such a large differential Li depletion between stars of the thick and thin discs.

Finally, in Fig. \ref{Fig:LiversusFeH_high} we also explore the issue of a {\it decreasing} Li abundance at supersolar metallicities, as suggested by the analysis of the AMBRE data. We find that such a decrease can be explained only under the assumption of a substantial reduction of the Li yields with metallicity: we have to assume that the adopted sources (2-5 \ms \ stars) have a negative net Li yield, decreasing strongly with metallicity above \zs.
We consider it difficult at this stage to conceive a plausible mechanism for such a decrease, such as, for example, absence of hot-bottom burning at high metallicities.  The yields of \cite{Karakas2016} suggest rather an increase of the Li yields as the metallicity increases from \zs \ to 2 \zs.

Alternatively, internal stellar depletion or 
some observational  artifact might explain the observed decline of Li/H. For instance, since the thin and thick disc 
sequences  merge at [Fe/H]$\ga$-0.2 in the [$\alpha$/Fe]-[Fe/H] plane,
it is difficult to disentangle those galactic components at high metallicity based on chemical tagging alone.
%As a consequence, due to biases in the target selection, 
Thus, the AMBRE sample at high metallicity could be contaminated by
metal-rich thick disc stars (although such  stars are not expected to be numerous in the thick disc).
Then, since thick disc stars are expected to be Li-poor, it could be that the Li depletion could
be artificially produced by an important amount of metal-rich thick disc stars.
Although such an observational bias is unlikely, it cannot be excluded. We suggest
therefore that the AMBRE Li depletion at super-solar metallicity
should be confirmed by further measurements in the future.

\section{Summary}
\label{sec:Summary}

In this work we investigate the behaviour of Li in the local disc, in light of recent observational data from AMBRE \citep{Guiglion2016}. Li/H appears to have a "two-branch" behaviour when plotted against Fe/H in the thin and thick discs, rminiscent of the well known behaviour of O/Fe versus Fe/H. We are using a semi-analytical model of disc evolution with parametrized radial migration, which has been successfully applied to the Milky Way disc and found to reproduce reasonably well most of its key observables \citep{Kubryk2015a}. In that model, the distinction between thin and the thick discs is assumed to be due to age, that is, thick-disc stars have ages $>$9 Gy.

We study the Li evolution in that model by considering both a low ("Spite plateau") and a high (SBBN) \lp \ value. We also consider different kinds of stellar Li sources, both long-lived (low-mass red giants of $\sim$1.5 \ms) and shorter-lived ones (2-5 \ms \ AGB stars). We include a detailed treatment of the GCR component of both Li isotopes.

We explain the "two-branch"  behaviour for  both O/Fe and Li/H as being due to the fact that most stars of the local thick disc originate in the inner Galaxy, where early evolution proceeded at a more rapid pace than in our vicinity. Those stars had little contamination from Fe from SNIa, thus displaying higher [O/Fe] ratios than thin disc stars of the same metallicity. For the same reason, thick disc stars have little contamination from stellar Li sources, when these are assumed to be long-lived stars: in that case, the Li abundance of thick disc stars reaches values only  slightly above the Spite plateau, while that of thin disc stars increases substantially with [Fe/H].

Our key finding is that our simple model of disc evolution with radial migration reproduces reasonably well the two-branch behaviour {\it assuming the stellar Li source is low-mass stars (or any long-lived source with appropriate Li yield) and \lp \ is the one of the Spite plateau}. Within these assumptions, no Li depletion is required in the envelopes of the most Li-rich low mass stars in that case in order to interpret the data. If  the high \lp \ value of SBBN is adopted instead, the two-branch behaviour is still found in the thin and thick discs;  however, that case leads to the necessity of assuming substantial minimal Li depletion $D_{min}$ in all stellar envelopes, which has to be a decreasing function of stellar metallicity (or age). 

We also show that, if the stellar Li source is assumed to be  AGB stars in the $\sim$2-5 \ms \ range there is no clear "two-branch" behaviour in the Li/H versus Fe/H plane, because the lifetimes of Li sources are then comparable to those of Fe sources. The Li versus Fe evolution differs little from one disc region to another and radial migration is of no help in creating differences in abundance patterns in that case.

Finally, we show how the puzzling feature revealed by AMBRE, namely the decreasing Li abundance at supersolar metallicities, requires the assumption of a strong reduction of the Li yields above solar metallicity. We cannot conceive at present such a mechanism on theoretical grounds. However, we cannot exclude internal stellar depletion or that the AMBRE finding is due to some systematic bias, for example, misidentifying thick disc stars as thin disc ones. Since this is the first time that the abundance of an element heavier than H is reported to decrease with metallicity, the AMBRE results should be  confirmed by further observations/analysis before such far reaching conclusions are drawn.  

\medskip

\noindent 
{\it Note added in proof:} A decline of Li/H at supersolar metallicities is also recently reported by the GES group (Fu et al. 2017, submitted), thus supporting the AMBRE result.
 
%\acknowledgements{We are grateful to the referee for suggestions that improved the presentation and refined the arguments of the text.} 

\bibliographystyle{aa} % style aa.bst
\bibliography{References} % your references Yourfile.bib

\end{document}